\documentclass[pra,twocolumn]{revtex4}

\usepackage{graphicx}
\usepackage{dcolumn}
\usepackage{bm}

\newcommand{\bo}{\raise-1mm\hbox{\Large$\Box$}}

\newcommand{\rr}{\mathbf{r}}
\newcommand{\kk}{\mathbf{k}}
\usepackage{epsfig}
\usepackage{latexsym}
\usepackage{mathrsfs}
\usepackage{setspace}
\usepackage{amsmath}
\usepackage{floatflt}
\usepackage{wrapfig}

\begin{document}

\title{Emergent structure in a dipolar Bose gas in a one-dimensional lattice}

\author{Ryan M. Wilson\email{rmw@colorado.edu}}
\author{John L. Bohn}

\affiliation{JILA and Department of Physics, University of Colorado, Boulder, Colorado 80309-0440, USA}

\date{\today}

\begin{abstract}
We consider an ultracold dipolar Bose gas in a one-dimensional lattice.  For a sufficiently large lattice recoil energy, such a system becomes a series of non-overlapping Bose-Einstein condensates that interact via the long-range dipole-dipole interaction  (ddi).  We model this system via a coupled set of non-local Gross-Pitaevskii equations (GPEs) for lattices of both infinite and finite extent.  We find significantly modified stability properties in the lattice due to the softening of a discrete roton-like mode, as well as ``islands'' in parameter space where biconcave densities are predicted to exist that only exist in the presence of the other condensates on the lattice.  We solve for the elementary excitations of the system to check the dynamical stability of these solutions and to uncover the nature of their collapse.  By solving a coupled set of GPEs exactly on a full numeric grid, we show that this emergent biconcave structure can be realized in a finite lattice with atomic $^{52}$Cr.
\end{abstract}

\maketitle

\section{Introduction}

Recent progress on the experimental realization of ultracold dipolar quantum fluids is providing an unprecedented opportunity to study these systems in detail.  Interesting dipolar effects have been demonstrated in systems with modest dipolar interactions, such as Bose-Einstein condensates (BECs) of $^{52}$Cr~\cite{LahayeNature,LahayeRev}.  Additionally, the achievement of a near-quantum-degenerate gas of fermionic $^{40}$K$^{87}$Rb~\cite{Ni08}, the transfer of bosonic $^{41}$K$^{87}$Rb into its rovibrational ground state~\cite{Aikawa10arXiv} and the cooling and trapping of atomic Dy~\cite{LuLev10} shows promise of a rich future for this field as these species host large dipole moments that may demonstrate new physics both in the mean-field regime and beyond.

Because of the anisotropic nature of the dipole-dipole interaction (ddi), the physics of a dipolar system depends strongly on the geometry of the trap in which it is held.  For example, inelastic scattering processes of both bosonic and fermionic species are predicted to be highly suppressed in tighter, quasi-two dimensional (q2D) traps when the trap is applied along the polarization axis of the dipoles~\cite{Quemener10arXiv,DIncao10arXiv,Micheli07,Quemener10,Micheli10}.  This suppression leads to more stable, longer-lived many-body systems of reactive species.  Additionally, it was shown using a BEC of atomic $^{52}$Cr that tighter confinement in the polarization direction energetically stabilizes a dipolar BEC (DBEC) against collapse~\cite{Koch08}.  So, tight trapping along the polarization axis is necessary to obtain stable, high density dipolar quantum fluids.

Such a trap is realizable in a one-dimensional (1D) optical lattice, where a laser is reflected onto itself and high and low intensity regions are formed by its interference pattern.  The presence of the lattice brings up an interesting point regarding the physics of such a system.  While the ddi is anisotropic, it is also long-range, scaling as $1/r^3$, and if the lattice spacing is sufficiently small then the effect of the ddi is non-negligable between the lattice sites.  For example, interlayer superfluidity is predicted to exist in two adjacent layers of polar fermions~\cite{Pikovski10arXiv}, and scattering in the 2D plane is predicted to be significantly modified by the presence of a weakly bound state of dipoles in adjacent layers~\cite{Klawunn10}.  Dramatic effects are predicted for layers of bosons, as well, for both q2D~\cite{Klawunn09,Wang08arXiv} and radially trapped~\cite{Koberle09,Junginger10} lattice sites.  The presence of the lattice is predicted to significantly alter the dispersion via the softening of a roton-like mode in the system, and thus to alter the stability properties of the Bose gas.

In this work, we consider a gas of bosonic dipoles in a 1D lattice with the dipoles polarized along the lattice axis, so that the system is cylindrically symmetric.  Assuming that the lattice recoil is sufficiently large, we model the potentials of the individual sites as cylindrically symmetric harmonic traps.  At ultracold temperatures, this leads to a lattice of non-overlapping DBECs coupled by the long-range ddi.  We study the stability of this system both for an infinite and finite 1D lattice.  Additionally, we find regions in parameter space where biconcave structure is predicted to exist that is emergent in the lattice system, in other words, that does not exist in a single condensate.  To ensure the accuracy of our results, we calculate the elementary excitations of the system and use them to determine whether our solutions are dynamically stable.  In doing so, we map the structure and stability of 1D lattice of purely dipolar DBECs.

\section{Formalism}

We consider an ultracold, dilute gas of bosonic dipoles in a 1D optical lattice in the $z$-direction with lattice spacing $d_\mathrm{lat}$.  If the lattice is sufficiently deep, it can be modeled by a series of $N_\mathrm{lat}$ harmonic traps, where each site is described by a cylindrically symmetric potential $U_j(\rr) = \frac{1}{2}M \omega_\rho^2 \left(\rho^2 + \lambda^2 (z-jd_\mathrm{lat})^2\right)$, where $M$ is the mass of the individual bosons and $\lambda=\omega_z / \omega_\rho$ is the trap aspect ratio.  This system is well-described by the coupled set of non-local Gross-Pitaevskii equations (GPEs)
\begin{equation}
\label{GPE}
\left\{ \hat{h}_j(\rr) + \sum_{j^\prime=1}^{N_\mathrm{lat}}\phi_d^{j^\prime}(\rr) - \mu_j \right\}\Psi_j(\rr) = 0,
\end{equation}
where $\hat{h}_j(\rr)$ is the non-interacting, or single-particle Hamiltonian
\begin{equation}
\hat{h}_j(\rr) = -\frac{\hbar^2}{2M}\nabla^2 + U_j(\rr),
\end{equation}
$\Psi_j(\rr)$ is the condensate wavefunction at site $j$, $j$ is an integer and $\mu_j$ is the corresponding chemical potential.  Without the presence of the long-range dipole-dipole interaction (ddi), these $N_\mathrm{lat}$ equations would be independent.  The ddi couples the equations through the mean-field potentials $\phi_d^j(\rr)$, given by the convolution
\begin{equation}
\label{mfr}
\phi_d^{j}(\rr) = \int d\rr^\prime V_d(\rr-\rr^\prime) n_j(\rr^\prime)
\end{equation}
where $n_j(\rr)=|\Psi_j(\rr)|^2$ is the density of the condensate occupying the $j^\mathrm{th}$ site with norm $\int d\rr^\prime n_j(\rr^\prime) = N_j$, $N_j$ is the condensate number for site $j$ and $V_d(\rr-\rr^\prime)$ is the two-body ddi potential for dipoles polarized along $\hat{z}$, given by~\cite{Yi00}
\begin{equation}
\label{ddi}
V(\rr-\rr^\prime) =  d^2 \frac{1-3\cos^2{\theta_{\rr-\rr^\prime}}}{|\rr-\rr^\prime|^3},
\end{equation}
where $d$ is the dipole moment of the bosons and $\theta_{\rr-\rr^\prime}$ is the angle between $\rr-\rr^\prime$ and $\hat{z}$.  A description of the fully-condensed, stationary state of this system of dipolar Bose-Einstein condensates (DBECs) is then given by the set of solutions $\{\Psi_j(\rr),\mu_j\}$ that minimize the energy functional corresponding to Eq.~(\ref{GPE}), given by
\begin{eqnarray}
\label{Efun}
E[\{\Psi_j(\rr)\}] &=& \sum_{j} \int d\rr\, \Psi_j^\star(\rr) \left\{ \hat{h}_j(\rr) \right. \nonumber \\
&{}& + \left. \frac{1}{2}\sum_{j^\prime=1}^{N_\mathrm{lat}}\phi_d^{j^\prime}(\rr)\right\}\Psi_j(\rr).
\end{eqnarray}

Generally, a full description of a dilute BEC of interacting atoms includes contact interactions given by the pseudo-potential $V_c(\rr-\rr^\prime)=g \delta(\rr-\rr^\prime)$ where $g\propto a_s$ and $a_s$ is the $s$-wave scattering length of the atoms.  This interaction is short-range, and results in the mean-field potential $\phi_c^j(\rr) = g |\Psi_j(\rr)|^2$.  Modeling a system of non-overlapping BECs in a 1D lattice interacting only via contact interactions results in a set of uncoupled GPEs.  While the interplay of contact and ddi interactions is predicted to produce interesting effects~\cite{Lu10} that would likely be modified by the presence of the lattice, we set $a_s=0$ in this work to illuminate purely dipolar effects.  Because of its long-range nature, the ddi does not produce a simple mean-field like the contact interaction, and requires particular attention.

In practice, the dipolar mean-field is calculated in $k$-space to eliminate the problems associated with the divergence of the ddi in real-space.  To do this, the Fourier transforms of the densities $n_j(\rr)$ and the ddi $V_d(\rr-\rr^\prime)$ must be calculated.  Where $\mathcal{F}$ is the Fourier transform operator,
\begin{equation}
\tilde{n}_j(\kk) = \mathcal{F}\left[ n_j(\rr) \right] \equiv \int d\rr n_j(\rr) e^{-i\kk\cdot\rr}.
\end{equation}
Here, it will prove useful to define the shifted densities $\nu_j(\rr) = n_j(\rr_j)$ where $\rr_j = \{\rho,z-jd_\mathrm{lat}\}$, so that all $\nu_j(\rr)$ are formally centered about the origin.  Then, $\nu_j(\rr_{-j}) = n_j(\rr)$, and we can write $\tilde{n}_j(\kk) = \mathcal{F}[\nu_j(\rr_{-j})]$.  With some simple manipulation, this expression reduces to
\begin{equation}
\label{nkj}
\tilde{n}_j(\kk) = \mathcal{F}\left[\nu_j(\rr)\right] e^{ik_z d_\mathrm{lat} j}.
\end{equation}
So, the $k$-space density of the DBEC at site $j$ can be rewritten as the Fourier transform of $n_j(\rr)$ translated into the local set of coordinates, with an additional exponential term accounting for this spatial translation.  Now, by the convolution theorem, the mean-field contribution from the DBEC at site $j$ can be written as
\begin{equation}
\label{dmf}
\phi_d^j(\rr) = \mathcal{F}^{-1}\left[  \tilde{V}_d(\kk) \tilde{n}_j(\kk) \right]
\end{equation}
where  $\tilde{V}_d(\kk)$ is the Fourier transform of the ddi~\cite{Goral02},
\begin{equation}
\label{Vdk}
\tilde{V}_d(\kk) = \frac{4\pi \hbar^2 a_{dd}}{M}\left( 3\frac{k_z^2}{k^2}-1 \right)
\end{equation}
and $a_{dd}=\frac{M d^2}{3\hbar^2}$ is the characteristic dipole length.  In this work, we calculate $\nu_j(\rr)$ directly by calculating the shifted condensate wavefunctions $\Phi_j(\rr)$ such that $\nu_j(\rr) = |\Phi_j(\rr)|^2$ and account for the spatial separation of the DBECs, or the presence of the lattice, with the expression given in Eq.~(\ref{nkj}).  So, the wavefunctions $\Psi_j(\rr)$ and $\Phi_j(\rr)$ are related by $\Phi_j(\rr) = \Psi_j(\rr_j)$.

\section{Wavefunction Ansatz}

For a single DBEC, calculating the mean-field energy on a full numeric grid has proven fruitful~\cite{Wilson09b,Lahaye09}, however, this method is very computationally expensive when considering multiple interacting DBECs, both in real- and $k$-space.  In real-space, the convolution integral for the dipole-dipole mean-field must be done directly, where there is no $1/r^3$ divergence if the condensates do not overlap.  In $k$-space, the grid must be large enough to resolve the entire lattice because of the $e^{ik_z d_\mathrm{lat} j}$ dependence of the $k$-space densities.  To avoid these problems, we consider solutions of the form $\Phi_j(\rr) = \psi_j(\rho)\chi_j(z)$ where
\begin{equation}
\label{chi}
\chi_j(z) = \frac{1}{\sqrt{1+A_{2,j}^2}}\left( \chi_{0,j}(z) + A_{2,j}\chi_{2,j}(z) \right),
\end{equation}
\begin{equation}
\label{chi0}
\chi_{0,j}(z) = \frac{1}{\sqrt{l_{z,j}}\pi^\frac{1}{4}} \exp{\left[ -\frac{z^2}{2 {l_{z,j}}^2} \right]}
\end{equation}
and
\begin{equation}
\label{chi2}
\chi_{2,j}(z) = \frac{1}{2\sqrt{2 l_{z,j}}\pi^\frac{1}{4}}\exp{\left[ -\frac{z^2}{2{l_{z,j}}^2} \right]}H_2\left(\frac{z}{{l_{z,j}}}\right),
\end{equation}
where $H_2(x)=4x^2-2$ is the second Hermite polynomial.  This ansatz includes the zeroth and second harmonic oscillator wavefunctions with variable width and relative amplitude.  Plugging this ansatz into the GPE and integrating out the $z$-dependence results in a modified GPE in the radial coordinate $\rho$ that also depends on the widths $l_{z,j}$ and the relative amplitudes $A_{2,j}$ of the axial wavefunctions, but not the $z$-coordinate explicitly.  We derive this modified GPE for a single DBEC, given by Eq.~(\ref{mGPE}), in Appendix~\ref{app:mGPE}.

To test the ansatz given in Eq.~(\ref{chi}), we apply it to the well known system of a single DBEC in a harmonic trap.  This system was predicted to exhibit, for certain trap geometries and ddi strengths, biconcave structures where the maximum density of the DBEC exists not in the center of the trap, but in a ring about the center of the trap~\cite{Ronen07}.  For example, such structure is predicted to exist in a trap with aspect ratio $\lambda=7$ for ddi strengths near the stability threshold.

\begin{figure}
\includegraphics[width=1.0\columnwidth]{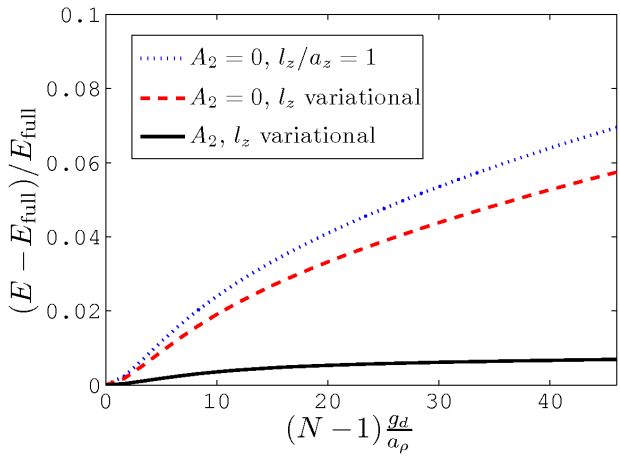}
\caption{\label{fig:El7} (color online)  The differences in energy of a DBEC in a trap with aspect ratio $\lambda=7$ as a function of interaction strength between that calculated using the ansatz given in Eq.~(\ref{chi}) and that calculated exactly on a full numeric grid.  The blue dotted line shows the energy difference calculated using no 2$^\mathrm{nd}$ order h.o. wavefunction and a fixed axial width $l_z=a_z$, the red dashed line shows the energy difference using the same wavefunction but with $l_z$ treated variational and the black solid line shows the energy difference including the 2$^\mathrm{nd}$ order h.o. wavefunction where the relative amplitude $A_2$ and $l_z$ are treated variationally.}
\end{figure}
\begin{figure}
\includegraphics[width=1.0\columnwidth]{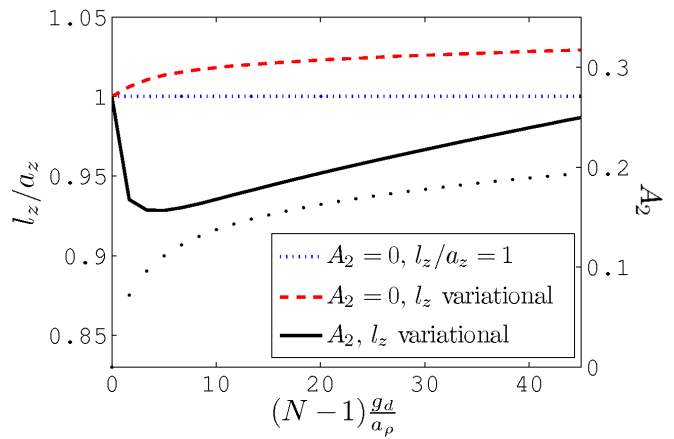}
\caption{\label{fig:gveccomp} (color online)  The values of the axial wavefunction parameters that, together with the radial wavefunction calculated on a grid, minimize the energy of a single DBEC in a trap with aspect ratio $\lambda=7$.  The blue dotted line shows the result for $l_z=a_z$, the red dotted line shows the values of $l_z$ when it is treated variationally and $A_2=0$ and the black solid line shows the values of $l_z$ when it and $A_2$, the black dotted line (marked by the right vertical axis) are both treated variationally.}
\end{figure}

Figure~\ref{fig:El7} compares the total energies of a single DBEC in a trap with $\lambda=7$ as a function of the ddi strength $(N-1)g_d/a_\rho$, where $a_\rho = \sqrt{\hbar / M \omega_\rho}$ is the radial harmonic oscillator length and $g_d = \frac{2\sqrt{2\pi}\hbar^2 a_{dd}}{M}$ is the ddi coupling, for various restrictions placed on the variational parameters of the axial wavefunction.  Plotted is the energy difference $(E-E_\mathrm{full})/E_\mathrm{full}$, where $E_\mathrm{full}$ is the energy calculated by solving the GPE exactly (within strict numerical precision) on a full numeric grid in $\rho$ and $z$.  The blue dotted line shows the energy of the DBEC when $A_2=0$ and $l_z$ is fixed to be the axial harmonic oscillator length, $a_z = \sqrt{\hbar/M \omega_z}$, the red dashed line shows the energy when $A_2=0$ and $l_z$ is treated variationally, and the black line shows the energy when $A_2$ and $l_z$ are both treated variationally.  Clearly, the full variational treatment is much more accurate than the cases where the second harmonic oscillator wavefunction is not included ($A_2=0$).  Indeed, it stays within 1\% of the exact energy for all values of $(N-1)g_d/a_\rho$ for which the DBEC is stable.  We find this to hold true for larger trap aspect ratios, as well.  Figure~\ref{fig:gveccomp} shows the values of the variational parameters for the same cases as in figure~\ref{fig:El7}.  In this figure, the left vertical axis labels $l_z/a_z$ and the left vertical axis labels $A_2$, shown by the black dots.

\begin{figure}
\includegraphics[width=1.0\columnwidth]{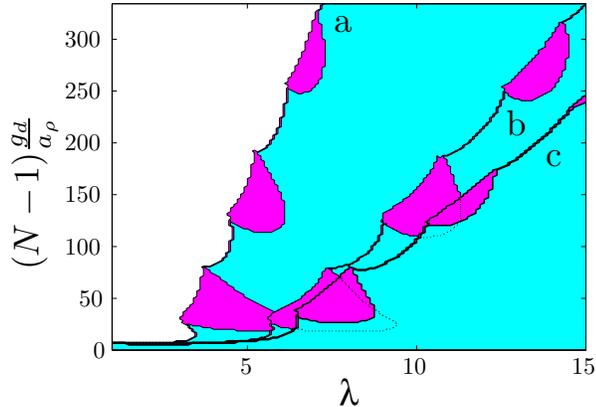}
\caption{\label{fig:stab} (color online) Structure/stability diagram for a single DBEC.  The colored regions indicate a dynamically stable condensate, and the pink (darker) regions indicate parameters for which the DBEC has biconcave density.  (a) and (b) are calculated using the ansatz for the axial wavefunction given in Eq.~(\ref{chi}) and (c) is calculated using a full numeric grid.  For (a), $A_2=0$ and $l_z/a_z=1$, and for (b), $A_2$ and $l_z$ are treated variationally.   }
\end{figure}

Beyond energetics, this ansatz also predicts semi-quantitatively the structure and stability of a single DBEC.  An interesting feature of the biconcave structure predicted in Ref.~\cite{Ronen07} is that it exists in ``islands'' of parameter space, defined by $(N-1)g_d/a_\rho$ and $\lambda$.  Figure~\ref{fig:stab} shows this structure/stability diagram for a single DBEC calculated using a) $A_2=0$ and $l_z/a_z=1$, b) $A_2$ and $l_z$ variational and c) a full numeric grid in $\rho$ and $z$.  Interestingly, the biconcave islands are present in each diagram and occur for almost exactly the same values of $(N-1)g_d/a_\rho$.  They are, however, shifted in $\lambda$, moving to smaller values as more restrictions are placed on the condensate wavefunction.  The diagram for the full variational ansatz (b) qualitatively matches that of the full numeric grid.  Thus, we expect that this ansatz will give physically meaningful results, if not quite quantitative ones. 

As we discuss below, a key benefit of this ansatz for the 1D lattice system is that it is analytic in $z$.  Another such ansatz that has this property is that of correlated Gaussians, which have been shown to reproduce the results of full numeric calculations for DBECs quite well~\cite{Rau10}.  However, we applied this ansatz to the lattice system and found that it is numerically unstable with the minimization techniques used here.

\section{Infinite lattice}
\label{inflat}

With confidence in the ansatz given in Eq.~(\ref{chi}), we now apply it to the 1D lattice system.  An interesting example to consider is that of an infinite lattice, with $N_\mathrm{lat}\rightarrow \infty$.  This approximation introduces a discrete invariance to the system so that we can set $\Phi_j(\rr) = \Phi_{j^\prime}(\rr)$ for all $j,j^\prime$.  Thus, we can neglect, for the time being, the indexing of the wavefunctions and let $\Phi_j(\rr) \rightarrow \Phi(\rr)$ for all $j$.  Then, the mean field potential at any site is, from Eq.~(\ref{nkj}), given by
\begin{equation}
\label{mfinf}
\phi_d(\rr) =  \sum_{j=-\infty}^{\infty} \int d\kk \tilde{V}_d(\kk) \tilde{n}(\kk)e^{ik_z d_\mathrm{lat} j} e^{-i\kk\cdot\rr}
\end{equation} 
where
\begin{eqnarray}
\label{nkhermite}
\tilde{n}(\kk) = \tilde{n}_\rho(k_\rho)\frac{e^{-\frac{1}{4}k_z^2 l_z^2}}{1+A_2^2} \nonumber \\
\times \left( 1+A_2\left( A_2-\left( \frac{1}{\sqrt{2}} + A_2\right)k_z^2 l_z^2 + \frac{1}{8}A_2 k_z^4 l_z^4 \right) \right).
\end{eqnarray}
We can manipulate the infinite sum in Eq.~(\ref{mfinf}) to give~\cite{Jeng00}
\begin{equation}
\label{comb}
\sum_{j=-\infty}^{\infty}e^{ik_z d_\mathrm{lat} j} = \frac{2\pi}{d_\mathrm{lat}}\sum_{j=-\infty}^{\infty} \delta\left( k_z -  \frac{2\pi j}{d_\mathrm{lat}}\right). 
\end{equation}
The term that accounts for the infinite lattice can therefore be written as a Dirac comb in $k_z$ with spacing $2\pi / d_\mathrm{lat}$ between peaks.  Inserting this expression into Eq.~(\ref{mfinf}) gives the mean-field potential 
\begin{equation}
\label{mfinf2}
\phi_d(\rr) = 2\frac{g_d}{d_\mathrm{lat}}\mathcal{F}_\mathrm{2D}^{-1}\left[ F_\mathrm{inf}(k_\rho)\tilde{n}_\rho(k_\rho) \right],
\end{equation}
where $2 g_d F_\mathrm{inf}(k_\rho)/d_\mathrm{lat}$ is the effective $k$-space ddi for the infinite lattice and $F_\mathrm{inf}(k_\rho)$ is given by
\begin{widetext}
\begin{equation}
\label{Finf}
F_\mathrm{inf}(k_\rho) = \sqrt{\frac{\pi}{2}}\sum_{j=-\infty}^{\infty}\frac{e^{-2\pi^2 j^2 \frac{l_z^2}{d_\mathrm{lat}^2}}}{(1+A_2^2)^2} \left[ 1+A_2\left( A_2-\left( \frac{1}{\sqrt{2}} + A_2\right)4\pi^2 j^2 \frac{l_z^2}{d_\mathrm{lat}^2} + 2 A_2\pi^4 j^4 \frac{l_z^4}{d_\mathrm{lat}^4} \right) \right]^2 \left( \frac{12\pi^2 j^2}{k_\rho^2 d_\mathrm{lat}^2 + 4\pi^2 j^2}-1 \right).
\end{equation}
\end{widetext}
The GPE for an infinite lattice of interacting DBECs is reduced to a single GPE in the radial coordinate $\rho$ where all of the axial dependence of the wavefunction is captured by the variational parameters $A_2$ and $l_z$.

We study the structure and stability of this infinite lattice of interacting DBECs by solving the modified GPE for the system (applying conjugate gradients~\cite{Ronen06} to minimize the corresponding energy functional) and studying the Bogoliubov de Gennes (BdG) excitations.  We find that the sum in Eq.~(\ref{Finf}) is sufficiently converged if a cutoff $j_\mathrm{cut}$ is applied to the index $j$ such that $j_\mathrm{cut}\gg d_\mathrm{lat}/2\pi l_z$.

Consistent with other results~\cite{Klawunn09,Wang08arXiv}, we find that the presence of the lattice serves to destabilize the system due to the softening of a discrete roton-like mode in the system.  For a single DBEC in a trap, tight axial confinement aligns the dipoles so that they are predominately repulsive and, for sufficiently low densities or interactions strengths, stabilizes the condensate.  In the presence of a 1D lattice, the attraction from the dipoles at other lattice sites extends the condensate in the axial direction, increasing the integrated axial density and, ultimately, making the system less stable.  This destabilization is made less dramatic as $d_\mathrm{lat}$ is increased.  

\begin{figure}
\includegraphics[width=1.0\columnwidth]{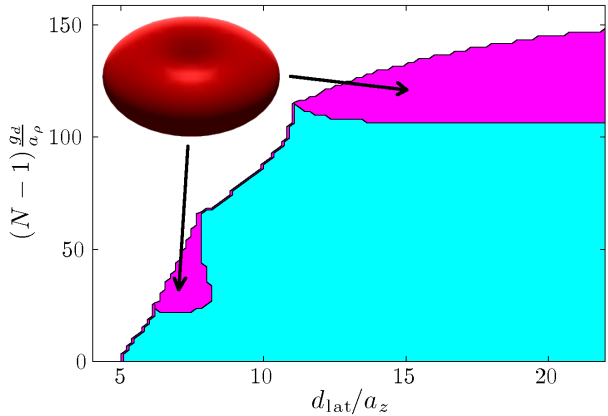}
\caption{\label{fig:l10} (color online)  Structure/stability diagram for an infinite lattice of DBECs in traps with aspect ratio $\lambda=10$ as a function of lattice spacing $d_\mathrm{lat}/a_z$ and interaction strength $(N-1)g_d/a_\rho$.  The colored region indicates dynamic stability, while the pink (darker) regions indicate parameters where the DBECs have biconcave density.  The inset shows an isodensity plot of a DBEC with biconcave density.}
\end{figure}

To study the structure and stability of the infinite lattice, we choose specific trap aspect ratios and explore the parameter space defined by $(N-1)g_d/a_\rho$ and $d_\mathrm{lat}$.  Figure~\ref{fig:l10} shows the region of dynamic stability for an infinite lattice of DBECs in traps with $\lambda=10$.  For lattice spacings $d_\mathrm{lat}/a_z \lesssim 5$, the condensate wavefunctions at adjacent sites overlap and the strong dipole-dipole attraction leads to complete instability.  In this figure, the colored regions indicate dynamic stability and the pink (dark) regions indicate parameters at which the DBECs exhibit biconcave density.  As $d_\mathrm{lat}/a_z$ is increased, the diagram approaches that given by a line at $\lambda=10$ in figure~\ref{fig:stab} for a single DBEC.  However, for smaller lattice spacings, a second biconcave island appears.  Without the presence of the lattice, biconcave structure would not exist for these parameters.  Thus, this structure is ``emergent'' in the lattice system.  The inset in figure~\ref{fig:l10} shows an isodensity plot of a DBEC with biconcave density.

\begin{figure}
\includegraphics[width=1.0\columnwidth]{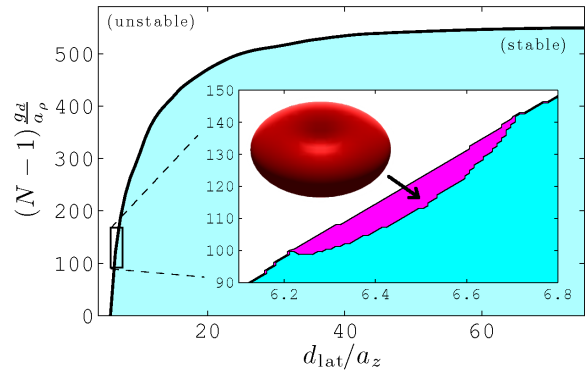}
\caption{\label{fig:l20} (color online)  Structure/stability diagram for an infinite lattice of DBECs in traps with aspect ratio $\lambda=20$ as a function of lattice spacing $d_\mathrm{lat}/a_z$ and interaction strength $(N-1)g_d/a_\rho$.  The inset shows a close-up of the diagram at the parameters indicated.  The pink (darker) region in the inset indicated parameters where the DBECs have biconcave density.  An isodensity plot of a DBEC with biconcave density is shown in this inset.}
\end{figure}

\begin{figure}
\includegraphics[width=1.0\columnwidth]{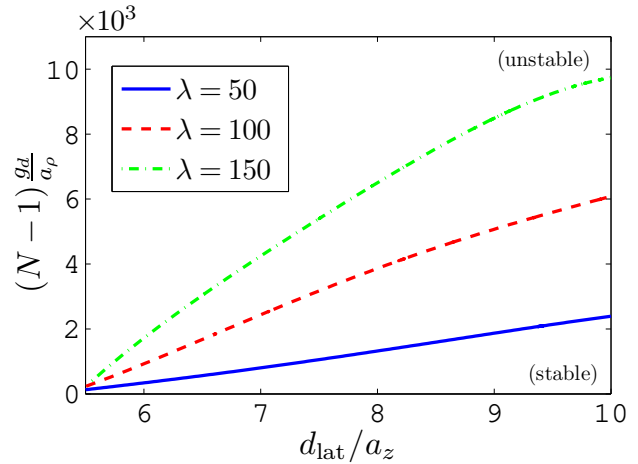}
\caption{\label{fig:biglambda} (color online)  Stability lines for an infinite lattice of DBECs for aspect ratios $\lambda=50,100,150$ as a function of lattice spacing $d_\mathrm{lat}/a_z$ and interaction strength $(N-1)g_d/a_\rho$.  The parameters beneath the lines are dynamically stable, while those above the lines are dynamically unstable.}
\end{figure}

Figure~\ref{fig:l20} shows the region of dynamic stability up to lattice spacings of $d_\mathrm{lat}/a_z=80$ for a infinite lattice of DBECs in harmonic traps with $\lambda=20$.  Here, the convergence of the stability line to $(N-1)g_d/a_\rho\sim 550$ is clear.  The inset shows a close-up view of the diagram where a biconcave island is predicted to exist.  As the aspect ratio is increased, the values of interaction strength $(N-1)g_d/a_\rho$ that the biconcave islands span becomes relatively smaller compared to the asymptotic value of the stability line.  Figure~\ref{fig:biglambda} shows the stability lines for infinite lattices with aspect ratios $\lambda=50,100,150$.  We find stability islands that exist at the stability threshold within the lattice spacings $d_\mathrm{lat}/a_z=6$ to $10$ for all of these aspect ratios.  Because they are so narrow in $(N-1)g_d/a_\rho$, though, they are not included in this plot.

By working in the $\rho$- and $z$-coordinates, a cylindrical symmetry is assumed.  However, it was shown in~\cite{Ronen07} that DBECs with biconcave densities are dynamically unstable to angular modes, or quasiparticles.  While the method used here is sensitive to dynamic instabilities that are purely radial, an extra step must be taken to detect angular instabilities.  

\subsection*{Bogoliubov de Gennes Equations}
\label{BdG}

The Bogoliubov de Gennes (BdG) equations describe the low-lying quasiparticles of the condensate.  They are derived by inserting the ansatz
\begin{equation}
\label{BdGansatz}
\psi(\rho) \rightarrow \left[ \psi(\rho) + \delta u(\rho)e^{i(m\varphi - \omega t)} + \delta v^\star(\rho)e^{-i(m\varphi-\omega t)}  \right]e^{-i\mu t}
\end{equation}
into the time-dependent GPE and linearizing about $\delta$, assuming that $\delta \ll 1$.  Here, $m$ is the quantum number describing the projection of angular momentum of the quasiparticle onto the $z$-axis.  In general, the energy eigenvalues of the BdG modes $\{ u,v^\star \}$ can be written as $\omega = \omega_R + i \omega_I$, where $\omega_R$ and $\omega_I$ are purely real.  When all $\omega_I=0$, the system is dynamically stable.  However, when some $\omega_I \neq 0$, the system is dynamically unstable, and the quasiparticle amplitude grows exponentially in time on a time scale $\sim 1/\omega_I$.

Like the single DBEC in~\cite{Ronen07}, we find that the biconcave structures in the infinite lattice are, for some critical density or ddi, dynamically unstable to angular quasiparticles with $m\geq 2$.  Ref.~\cite{Wilson09b} shows that this angular instability leads to angular collapse, or collapse with angular nodes, of the biconcave DBECs.  A measurement of the character of collapse, whether it be radial or angular, then provides a tool to map the structure along the stability threshold of the system.  

In our analysis, we found exotic ground state densities very close to the stability threshold, like those found for a finite lattice in Ref.~\cite{Koberle09}.  These solutions host multiple radial density oscillations, however, we find they are dynamically unstable and are thus unlikely to be experimentally observable.  

\section{Finite Lattice}
\label{finlat}

While the infinite lattice of DBECs provides a clear, simple example of emergent structure in this system, it is a difficult system to realize experimentally.  In a realistic experiment, the lattice has a finite extent and the occupations of the sites vary from site to site.  To model this more realistic lattice system, we consider an odd number of occupied lattice sites indexed by $j\in [-j_\mathrm{lat},j_\mathrm{lat}]$ where $j_\mathrm{lat} = (N_\mathrm{lat}-1)/2$ with particle number given by a Gaussian distribution, $N_j = N_\mathrm{max} \exp{[-(j/j_\mathrm{lat})^2]}$, where $N_\mathrm{max}$ is the particle number in the condensate in the center of the lattice at site $j=0$, and the outer-most sites have particle number $N_\mathrm{max}/e$~\cite{Koberle09}.

Instead of using an analytic form for the axial parts of the condensate wavefunctions, we solve the coupled set of GPEs given by Eq.~(\ref{GPE}) on a full grid (large enough to encapsulate the entire lattice) in $\rho$ and $z$ for each condensate.  We find good convergence by using the conjugate gradients method to minimize the full energy functional of the system~\cite{Ronen06}.  Additionally, to ensure numerical precision we apply a cutoff to the ddi in $\rho$ and $z$ so that a relatively small grid can be used while eliminating the effects of artificial ``image'' condensates that are present due to the use of the FFT algorithm in our calculation~\cite{Lu10}.

\begin{figure}
\includegraphics[width=1.0\columnwidth]{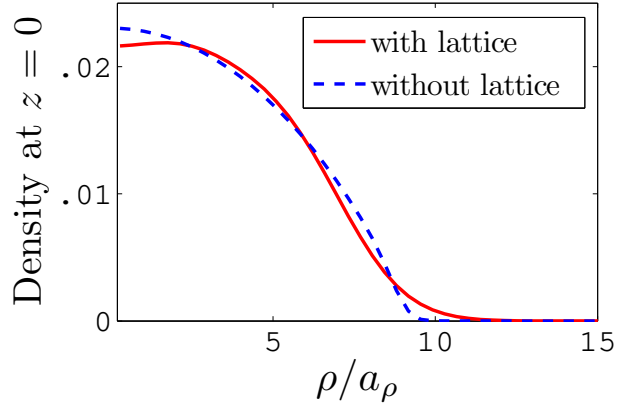}
\caption{\label{fig:latbicon} (color online)  Radial densities at $z=0$ of a DBEC with $(N-1)g_d/a_\rho = 550$ in a trap with aspect ratio $\lambda=50$.  The blue dashed line shows the density of a DBEC in a single harmonic trap, and the red solid line shows the density of a DBEC in the center site of a 1D lattice with nine occupied sites ($j_\mathrm{lat} = 4$).   This DBEC exhibits biconcave structure, while the single DBEC does not, demonstrating the emergence of this structure in the lattice system.  These densities were calculated by solving the GPE (coupled GPEs) exactly on a full numeric grid.}
\end{figure}

As an example, we consider a lattice with trap aspect ratios $\lambda=50$, $j_\mathrm{lat}=4$ (corresponding to $9$ occupied lattice sites), lattice spacing $d_\mathrm{lat} = 8 a_z$ and $(N_\mathrm{max}-1)g_d/a_\rho = 550$ on a numeric grid of size $[N_z,N_\rho] = [1024,128]$.  Figure~\ref{fig:latbicon} shows the density at $z=0$ of a DBEC at the center of the lattice ($j=0$) and, for comparison, the density of a DBEC with the same trap aspect ratio and ddi strength $(N-1)g_d/a_\rho=550$ but without the presence of the lattice.  While the DBEC in the single trap does not exhibit biconcave structure, the DBEC in the lattice does, showing that this emergent structure in the lattice system is present not only in the infinite lattice system, but also in the experimentally realistic system of a finite lattice with variable occupancy.  Indeed, such a system is realizable with atomic $^{52}$Cr, having a permanent magnetic dipole moment of $\mu = 6 \mu_B$ where $\mu_B$ is the Bohr magneton, axial harmonic oscillator frequencies of $\omega_z = 2\pi \times 30$ kHz and a maximum condensate occupancy of $N_\mathrm{max} \simeq 77\times 10^3$ atoms.

\section{Conclusion}
\label{conclusion}

In conclusion, we have mapped the structure and stability of a lattice of interacting, purely dipolar DBECs.  By asserting an analytic form for the axial part of the condensate wavefunctions (Eqs.~(\ref{chi})-(\ref{chi2})), we derive a simple, modified GPE for the radial part of the wavefunctions when the lattice is infinite.  We find isolated regions (``islands'') in the parameter space defined by the lattice spacing and the ddi strength where the DBECs are predicted to exhibit biconcave densities, where the maximum density exists not in the center of the trap but in a ring about the center of the trap.  To model a more experimentally realistic system, we consider a finite lattice with varying condensate number and solve the coupled set of GPEs exactly on a full numeric grid.  In doing so, we show that this emergent biconcave structure should be observable in a finite 1D lattice of DBECs of atomic $^{52}$Cr.

\begin{acknowledgments}
We acknowledge the financial support of the NSF and the DOE, and would like to thank Juliette Billy and Jonas Metz for their enlightening discussions.
\end{acknowledgments}

\appendix

\section{Modified GPE using $0^\mathrm{th}$ and $2^\mathrm{nd}$ harmonic oscillator wavefunctions}
\label{app:mGPE}

Consider the ansatz given by Eqs.~(\ref{chi})-(\ref{chi2}) for a system with a single harmonically trapped DBEC, so the indexing of the condensate wavefunction can be ignored and we can simply write $\Phi(\rr) = \psi(\rho)\chi(z)$.  We derive the modified GPE by multiplying the (dimensionless) GPE,
\begin{equation}
\left\{ -\frac{1}{2}\nabla^2 + U(\rr) + \phi_d(\rr) -\mu  \right\} \Phi(\rr)=0,
\end{equation}
by $\chi(z)$ and integrating over $z$.  This operation gives the modified GPE,
\begin{equation}
\label{mGPE}
\left\{ \hat{h}_\mathrm{eff}(\rho) + 2\frac{g_d}{l_z} \mathcal{F}^{-1}_\mathrm{2D}\left[ \tilde{n}_\rho(k_\rho) F_\mathrm{eff}\left( \frac{k_\rho l_z}{\sqrt{2}} \right)\right] \right\} \psi(\rho) = 0,
\end{equation}
where $\hat{h}_\mathrm{eff}(\rho)$ is the effective single-particle Hamiltonian,
\begin{eqnarray}
\label{h}
\hat{h}_\mathrm{eff}(\rho) = -\frac{1}{2}\nabla_\rho^2 + \frac{1}{2}\omega_\rho^2 \rho^2 - \mu + \frac{1}{1+A_2^2} \nonumber \\
\times \left[ \left( \frac{1}{l_z^2}+\lambda^2 l_z^2 \right)  \left( \frac{1}{4}+\frac{5}{4}A_2^2 \right) -\frac{A_2}{\sqrt{2}} \left(\frac{1}{l_z^2} - \lambda^2 l_z^2 \right) \right],
\end{eqnarray}
and $F_\mathrm{eff}(x)$ is given by
\begin{widetext}
\begin{eqnarray}
 F_\mathrm{eff}(x) = \frac{1}{\left(1 +  A_2^2 \right)^2}\left(1 + \sqrt{2} A_2\left(3 x^2 -1 \right) + \frac{3}{4} A_2^2\left(3 + 5 x^2 + 6 x^4 \right) + \frac{1}{4\sqrt{2}} A_2^3\left(1 + 9 x^2 + 42 x^4 + 12 x^6 \right) \right. \nonumber \\
+ \left. \frac{1}{64} A_2^4\left(41 + 3 x^2\left(81 + 134 x^2 + 60 x^4 +  8 x^6 \right) \right) - \frac{3\sqrt{\pi}}{2} \left[1 +  A_2\left( A_2\left(2 +  A_2^2 \right) \right. \right. \right. \nonumber \\
+ \left. \left. \left. 2 \left(\sqrt {2} + 2 A_2 \right)\left(1 +  A_2^2 \right) x^2 + A_2\left(3 + 4\sqrt {2} A_2 + 5 A_2^2 \right) x^4 +  A_2^2\left(\sqrt {2} + 2 A_2 \right) x^6 +  \frac{1}{4} A_2^3 x^8 \right) \right] x e^{x^2}\mathrm{Erfc}\left[ x \right] \right)
\end{eqnarray}
\end{widetext}
and $\mathrm{Erfc}[x]$ is the complimentary error function.  The corresponding mean-field energy due to the ddi is then
\begin{equation}
E_d = \frac{g_d}{l_z} \int d^2\rho\, n_\rho(\rho)\mathcal{F}_\mathrm{2D}^{-1}\left[ \tilde{n}_\rho(k_\rho)F_\mathrm{eff}\left( \frac{k_\rho l_z}{\sqrt{2}} \right)\right].
\end{equation}

\end{document}